\documentclass[aps,prl,twocolumn,fleqn,superscriptaddress]{revtex4}

\usepackage{graphicx,epsfig}
\usepackage{amsmath,amssymb,amsfonts}

\DeclareMathOperator{\tr}{Tr}

\begin{document}

\title{Near--Fields and Initial Energy Density
in High Energy Nuclear Collisions}

\author{R.~J.~Fries}
\affiliation{School of Physics and Astronomy, University of Minnesota,
  Minneapolis, MN 55455}
\affiliation{Cyclotron Institute and Department of Physics, Texas A\&M
  University, College Station, TX 77843}
\affiliation{RIKEN BNL Research Center, Brookhaven National Laboratory,
  Upton, NY 11973}

\author{J.~I.~Kapusta}
\affiliation{School of Physics and Astronomy, University of Minnesota,
Minneapolis, MN 55455}

\author{Y.~Li}
\affiliation{School of Physics and Astronomy, University of
Minnesota, Minneapolis, MN 55455}
\affiliation{Department of Physics and Astronomy, Iowa State University,
Ames, IA 50011}

\date{\today}

\begin{abstract}
We calculate the classical gluon field created at early times in collisions
of large nuclei at high energies. We find that the field is dominated by the
longitudinal chromoelectric and chromomagnetic components. We estimate the
initial energy density of this gluon field to be approximately
260 GeV/fm$^3$ at RHIC.
\end{abstract}

\maketitle

Experiments are being carried out at the Relativistic Heavy Ion Collider 
(RHIC) and soon will be at the CERN Large Hadron Collider (LHC) to create and
study quark gluon plasma.
Data from RHIC indicate that in collisions of gold nuclei at $\sqrt{s_{NN}}
= 200$ GeV energy densities far in excess of the critical value required for
deconfinement ($\epsilon_c \approx 2$ GeV/fm$^3$) are reached \cite{rhic-wp}.
Furthermore, the partonic phase seems to be thermalized after a very short
time $\tau_0<1$ fm/$c$. While the evolution of the quark gluon plasma in
equilibrium can be described by relativistic hydrodynamics \cite{KH}, the
initial soft interactions of the nuclei and the thermalization process before
the time $\tau_0$ are still not completely understood.

It has been argued that the initial dynamics for the collision of two
very high energy nuclei is determined by a universal phase called the
color glass condensate (CGC). This idea is based on gluon saturation
at a scale $~Q_s$ \cite{MV,KMW,JMKMLW:96,cgc,McL:05}.
Slowly evolving and randomly
distributed color charges in the nuclei are the sources of this gluon field.
A simple implementation is the McLerran-Venugopalan (MV) model \cite{MV,KMW}
in which the gluon field is given by the solution of the classical
Yang-Mills equations.

In this Letter we calculate the gluon field at early times after the collision
in the framework of the McLerran-Venugopalan model. We use an expansion of the
Yang-Mills equations in powers of the proper time $\tau$.
This is a near--field approximation which may be the most appropriate
use of the color glass condensate picture. We also estimate the initial energy
density at the time of overlap of the nuclei using a simple model for the
nuclear gluon distribution and coarse-graining methods to avoid ultraviolet
(UV) singularities. More details and a discussion of applications will be
provided elsewhere \cite{FKLMB}.

In high energy collisions the two colliding nuclei are highly Lorentz
contracted; therefore, the valence and large-$x$ partons are described
by infinitesimally thin sheets propagating on the light cone. Although
each nucleus is color neutral as a whole, local color fluctuations do
occur. At the moment of overlap, the color distributions in nucleus 1
($+$ light cone) and 2 ($-$ light cone) are $\rho_{1} (\mathbf{x}_\perp)$
and $\rho_{2} (\mathbf{x}_\perp)$, respectively. We use light cone
coordinates $x^{\pm} = \left(x^0\pm x^3\right)/ \sqrt{2}$ and
${\bf x}_{\perp} = (x^1,x^2)$. The distributions $\rho_k=\rho_k^a t^a$
($k=1,2$) are functions with values in $SU(3)$. Since they resemble
fluctuations of color we have to take the ensemble average of all
allowed functions $\rho_k$ at the end.
It is convenient to choose an axial gauge defined by $A^+ x^- + A^- x^+ = 0$.
In this gauge the current generated by the charges $\rho_k$ takes the form
$J^\pm = \delta(x^\mp)g\rho_{1,2} ({\bf x}_{\perp})$ and $J^i = 0$,
satisfying the equation of continuity $\left[ D_\mu, J^\mu\right]=0$.
The gluon field generated by this current can be obtained by solving the
Yang-Mills equations $[D_{\mu},F^{\mu\nu}]=J^{\nu}$.

The gauge potential $A^\mu$ is a smooth function of $x^{\mu}$ except for lines
with propagating charge. We follow the authors of ref.\ \cite{KMW} who showed
that the ansatz
\begin{eqnarray}
A^\pm(x) &=& \pm \theta(x^+)\theta(x^-)x^\pm\alpha(\tau,{\bf x}_{\perp}) \, ,\\
A^i(x) &=& \theta(x^-)\theta(-x^+)\alpha_1^i({\bf x}_{\perp}) + \theta(x^+)\theta(-x^-)\alpha_2^i({\bf x}_{\perp}) \nonumber \\
&+& \theta(x^+)\theta(x^-)\alpha_3^i(\tau,{\bf x}_{\perp}) \, .
\end{eqnarray}
satisfies the Yang-Mills equations in the different
sectors of Minkowski space.
Here upper Latin indices $i$, $j$, $\ldots$ always refer to transverse
components. The
$\alpha_1^i$ and $\alpha_2^i$ are the purely transverse gauge
potentials of nucleus 1 and 2, respectively. They can be written with
the help of transformation matrices $U_{k}=e^{i\phi_{k}}$, $k=1,2$,
such that they are gauge transformations of the vacuum:
\begin{align}
  \alpha_{k}^i({\bf x}_\perp) & = \frac{i}{g} U^{-1}_{k}
  \partial^i U_{k} \, \text{, with} \label{eq:bc0.1}  \\
  \nabla_\perp^2 \phi_{k}(\mathbf{x}_\perp) &= g^2 \rho_{k}
  (\mathbf{x}_\perp).
  \label{eq:bc0.2}
\end{align}

$\alpha$ and $\alpha_3^i$ describe the field in the forward light cone
($x^+ > 0$, $x^- > 0$) which is generated in the collision. They are
smooth functions of $\mathbf{x}_\perp$ and the proper time
$\tau=\sqrt{2x^+x^-}$. They are independent of the space-time rapidity
$\eta = 1/2 \ln (x^+/x^-)$ because the current $J^\mu$ is boost-invariant.
In the forward light cone the Yang-Mills equations
can be rephrased as \cite{KMW}:
\begin{eqnarray}
  \frac{1}{\tau^3}\partial_{\tau}\tau^3\partial_{\tau}\alpha-[D^i,[D^i,\alpha]]
  &= 0 \, , \label{eq:EOM1}\\
  \frac{1}{\tau}[D^i,\partial_{\tau}\alpha^i_3] -ig\tau[\alpha,\partial_{\tau}
  \alpha] &= 0 \, , \label{eq:EOM2}\\
  \frac{1}{\tau}\partial_{\tau}\tau\partial_{\tau}\alpha^i_3-ig\tau^2[\alpha,
  [D^i,\alpha]]-[D^j,F^{ji}] &= 0 \, . \label{eq:EOM3}
\end{eqnarray}
$\alpha$ and $\alpha_3^i$ are connected with the single nucleus fields
$\alpha^i_{1,2}$ via boundary conditions at $\tau=0$:
\begin{eqnarray}
  \alpha_3^i (\tau=0,{\bf x}_\perp) &=& \alpha_1^i ({\bf x}_\perp) + \alpha_2^i
  ({\bf x}_\perp) \, , \label{eq:bc_boost2}\\
  \alpha (\tau=0,{\bf x}_\perp) &=& -\frac{i g}{2} \left[
  \alpha_1^i({\bf x}_\perp),\alpha_2^i ({\bf x}_\perp) \right] \, .
  \label{eq:bc_boost3}
\end{eqnarray}
An explicit analytic solution of \eqref{eq:EOM1}--\eqref{eq:EOM3} is not
known. However, lowest order perturbative solutions \cite{KMW} as well 
as numerical solutions \cite{KV:98,Lappi:03} are available.

Decoherence and pair production \cite{GKT:05,KharTuch:05} will eventually 
destroy the classical field and lead to thermalization. Typical values for 
the thermalization time $\tau_0$ used in hydrodynamic calculations range 
from 0.15 fm/$c$ to 1.0 fm/$c$ \cite{rhic-wp,KH}.
It is clear that the classical description breaks down before $\tau_0$.
Hence, what we can hope to calculate in this particular framework is
the short-term behavior of the gluon field, i.e.\ the
near-field close to the light cone.
The functions $\alpha$ and $\alpha_3^i$ are regular at $\tau=0$.
Therefore it is legitimate to solve the Yang-Mills equations using
a power series in $\tau$. We write
\begin{equation}
  \alpha(\tau,{\bf x}_{\perp}) =
  \sum_{n=0}^{\infty}\tau^n\alpha_{(n)}({\bf x}_{\perp}) \, ,
  \label{eq:powers1}   
\end{equation}
and similarly for $\alpha_3^i$. We also use expansions for the
field strength tensor and covariant derivative in the forward light cone with
coefficients $F^{\mu\nu}_{(n)}$ and $D^\mu_{(n)}$, respectively.

Using these expansions in Eqs. \eqref{eq:EOM1} through
\eqref{eq:EOM3} yields an infinite set of equations for the coefficients
$\alpha_{(n)}$ and $\alpha_{3(n)}^i$. To lowest order in $\tau$, the
fields are just given by the boundary conditions
\eqref{eq:bc_boost2} and \eqref{eq:bc_boost3},
\begin{align}
  \alpha^i_{3(0)} & = \alpha_1^i + \alpha_2^i \, , \\
  \alpha_{(0)} & = -ig [\alpha_1^i,\alpha_2^i]/2 \, .
\end{align}
It is now possible to give a solution for any order in $\tau$
recursively. It is straightforward to prove that for $n>1$
\begin{align}
  \alpha_{(n)} &= \frac{1}{n(n+2)} \sum_{k+l+m=n-2} \left[ D^i_{(k)},
  [ D^i_{(l)},\alpha_{(m)}] \right]\, , \nonumber \\
  \alpha^i_{3(n)} &= \frac{1}{n^2}\left( \sum_{k+l=n-2}
  \left[ D^j_{(k)}, F^{ji}_{(l)} \right] \right. \\
    & \quad \left. + ig \sum_{k+l+m=n-4}
  \left[ \alpha_{(k)}, [ D^i_{(l)},\alpha_{(m)} ] \right] \right).
  \nonumber
\end{align}
One can immediately conclude that these fields vanish for all odd
powers of $\tau$: $\alpha_{(2n+1)}=0$, $\alpha_{3(2n+1)}^i=0$.
Similar recursion relations hold for the field strength tensor. For brevity
we only cite the relation for the longitudinal chromoelectric field
$E_z = F^{+-}$ which is
\begin{equation}
  F^{+-}_{(n)} = \sum_{k+l+m=n-2} \frac{1}{n(m+2)} \left[ D^i_{(k)},
  [ D^i_{(l)},F^{+-}_{(m)}] \right]\, .
\end{equation}
A summation of the recursive solution in closed form does not seem
feasible. However, we assert that an analysis using just the first 
few orders in $\tau$ is extremely useful.

The non-vanishing components of the field strength for the lowest three
orders in $\tau$ are
\begin{align}
  F^{+-}_{(0)} &= ig \left[ \alpha_1^i,
  \alpha_2^i \right] \, ,\\
  F^{21}_{(0)} &= ig \epsilon^{ij} \left[ \alpha_1^i,
  \alpha_2^j \right] \, , \\
  F^{i\pm}_{(1)} &= -\frac{e^{\pm\eta}}{2\sqrt{2}} \left(
  [ D^j_{(0)},F^{ji}_{(0)} ] \pm [ D^i_{(0)}, F^{+-}_{(0)} ] \right) \, , \\
  F^{+-}_{(2)} &=
  \frac{1}{4} [ D^i_{(0)} , [ D^i_{(0)} , F^{+-}_{(0)} ]] \, , \\
  F^{21}_{(2)} &= \frac{1}{4} [ D^i_{(0)} , [ D^i_{(0)} , F^{21}_{(0)} ]]
  \, .
\end{align}
Here $\epsilon^{ij}$ is the antisymmetric tensor. Thus the longitudinal
chromoelectric field $E_z$ and the longitudinal chromomagnetic field
$B_z = F^{21}$ start with finite values at $\tau = 0$.
The transverse electric and magnetic fields, which are linear
combinations of the components $F^{i\pm}$, are zero at $\tau = 0$ and
start at order $\tau^1$. Generally, longitudinal fields have only
contributions from even powers in $\tau$, transverse fields
only consist of odd powers in $\tau$.

This observation leads to the following space-time picture.
Inside the nuclei the color sources create purely transverse fields
$F^{i\pm} = \delta(x^\mp) \alpha_{1,2}^i$ on the
light cone. This is completely analogous to the abelian case.
After nuclear overlap, non-abelian interactions between these fields create
strong longitudinal chromoelectric and chromomagnetic fields, while the onset
of transverse fields in the forward light cone is delayed. The situation
resembles a capacitor with a longitudinal field, but it is important to
realize that only the non-abelian nature of the gluon field can generate such
a field for recoilless charges receding from each other with the speed of
light.
The strong longitudinal fields at early times are an immediate consequence of
the equations of motion; however, this fact has not received much attention
before. Recently the strong pulse of longitudinal fields
and its possible consequences have been discussed
\cite{FKL:05,KharTuch:05,KhaVen:01,LaMcL:06}.

In the second part of this Letter we would like to use our results to
discuss the initial energy density $\epsilon_0 = \langle T^{00}_{(0)}
\rangle$ for $\tau\to 0$. Here $T^{\mu\nu}$ is the energy momentum
tensor of the classical field which we expand in powers of $\tau$
as well. We postpone all further discussion to a later publication
\cite{FKLMB}. Note that the recursion formulas use the gluon fields
$\alpha^i_{k}$ as the starting point. However those
have to be determined by solving the Yang-Mills equations (\ref{eq:bc0.1}),
(\ref{eq:bc0.2}) for a single nucleus which is a difficult task.
For our discussion here, difficulties associated with
non-linearities in the boundary conditions are simplified by a mean-field
approximation. As we will argue below it still represents the essential
physics of the full solution. We achieve this by replacing Eqs.\
\eqref{eq:bc0.1} and \eqref{eq:bc0.2} with
\begin{equation}
  \left( \nabla^2_{\perp}  - \frac{1}{R_c^2} \right) \phi_k = g^2 \rho_k \, ,
  \quad \alpha_k^i = -\frac{1}{g}\partial^i \phi_k \, .
  \label{eq:approx}
\end{equation}
The solution is formally linear in $\rho_k$. However, we introduced a
screening length $R_c$ as a parameter which will depend on the
charge distribution $\rho_k$.

The idea behind this approximation to Eqs.\ (\ref{eq:bc0.1}) and 
(\ref{eq:bc0.2}) is as follows. To lowest order in the charge
density $\rho_k$ we have
$\alpha_k^i = -(1/g) \partial^i \phi_k + \mathcal{O}(\rho_k^3)$. The
primary effect of the non-linearities is a partial
screening of the field on length scales $\sim 1/Q_s$
\cite{JMKMLW:96}. However, the screening is incomplete
unless confinement is enforced in addition \cite{LamMa:99}.
In our approximation screening is provided by the scale $R_c \sim 1/Q_s$
and the result is perfectly infrared safe.
For a reasonable estimate of the screening effect one can invoke the
analogy to electric screening in a QCD plasma which implies
$R_c^{-2} = 4\alpha_s \sigma/3$ where $\sigma$ is the area density of the
number of color charges.

Let us now consider the two nuclei as being made up of two ensembles of 
discrete charges, given by $SU(3)$ matrices $T_{1,u}$ and $T_{2,u}$ at 
transverse positions $\mathbf{b}_{u}$.
The charge densities can be written as
\begin{equation}
  \rho_k (\mathbf{x}_\perp) = \sum_u R(|\mathbf{x}_\perp -
  \mathbf{b}_{u}|)  T_{k,u} \, .
\end{equation}
The index $u$ represents coarse--grained cells in both nuclei with
a certain number of color sources, $N_{1,u}$ and $N_{2,u}$, respectively,
in each of them, and $R(r)$ is the spatial distribution of the effective
color charge in each cell normalized to one. Note that the factorization
of spatial and color degrees of freedom makes sense if the coarse-grained cells
have sizes much smaller than $1/Q_s$.

Then from Eqs.\ \eqref{eq:bc0.1}, \eqref{eq:bc0.2} it follows that the
gauge potentials can be written as
\begin{equation}
  \alpha^i_k (\mathbf{x}_\perp) = g \sum_u
  \frac{x_\perp^i-b_{u}^i}{|\mathbf{x}_\perp - \mathbf{b}_{u}|} \tilde
  T_{k,u} 
  G(|\mathbf{x}_\perp - \mathbf{b}_{u}|)
  \label{eq:nucfield}
\end{equation}
where $G$ is the linear field of a single charge, i.e.\
$G(r) = d\varphi/dr$ with $\nabla^2_\perp \varphi(r) = R(r)$.
The $\tilde T_{k,u}$ are $SU(3)$ valued functions and can be
interpreted as the modifications of the charges $T_{k,u}$ through
non-linear interactions. One can expand $\tilde T_{k,u}(r) = T_{k,u}
+ \mathcal{O}(T^3)$, where the first term is the linear
contribution, and higher order terms reflect the screening from
interactions with charges in neighboring cells.
We can now simplify the situation by applying the approximation
\eqref{eq:approx}. It amounts to the replacement $\tilde T_{k,u} G \to T_{k,u}
\tilde G$. The screening present in $\tilde T$ is now factorized into
a modified field profile $\tilde G$ with
\begin{equation}
  \left( \nabla^2_{\perp}  - \frac{1}{R_c^2} \right) \tilde\varphi(r) 
  =  R(r) \, ,
  \quad \tilde G(r) = d\tilde\varphi/dr \, .
  \label{eq:approx_1}
\end{equation}

In addition, we have to impose an ultraviolet (UV) cutoff $Q_0$.
The McLerran-Venugopalan model does not provide a UV finite answer for the
energy density at $\tau = 0$, as also noticed in \cite{Lappi:06}. 
This comes from the fact that hard modes with momentum much larger than 
$Q_s$ are treated correctly. They are better described by hard 
perturbative processes. Therefore $Q_0$ should be chosen to be the 
cutoff between hard processes involving modes with transverse momentum 
$p_T > Q_0$, and the bulk modes with $p_T < Q_0$. We realize that our 
coarse-graining provides this cutoff. To be more precise we choose 
Gaussian profiles $R$ for each charge with a width $\lambda=1/Q_0$.
Putting everything together we find the modified field profile is approximately
\begin{equation}
  \tilde G(x_\perp) \approx \frac{1-\exp(-x_\perp^2/\lambda^2)}{2\pi R_c} K_1
  \left(\frac{x_\perp}{R_c} \right) \, ,
  \label{eq:finalg}
\end{equation}
where $K_1$ is a modified Bessel function.

To calculate the expectation values of observables
we discretize the functional integrals over the charge distributions
$\rho_k$ and replace them with integrals over the group $SU(3)$ at
each point $\mathbf{b}_u$.
The correct weight function to be used for the integral for cell $u$ in
nucleus $k$ is
$w_{N_{k,u}}(T_{k,u}) = (N_c/(\pi N_{k,u}))^4 \exp(-N_c T^2_{k,u}/N_{k,u})$
where $N_c=3$ \cite{JeVe}. As a straightforward
generalization of the results in \cite{JeVe} we use
$N_{k,u} = N^q_{k,u} + N^{\bar q}_{k,u} + C_A/C_F N^g_{k,u}$
where $N^q_{k,u}$, $N^{\bar q}_{k,u}$ and $N^g_{k,u}$ are the number
of quarks, antiquarks and gluons in each cell.
We define the area charge density in nucleus $k$ as
$\sigma_{k,u} = N_{k,u}/(\text{area of the cell } u)$ for each cell. It is
then straightforward to define a continuous charge density
$\sigma_k(\mathbf{x}_\perp)$. The saturation scale usually contains 
the strong coupling and we set $Q_s^2 = \alpha_s \sigma$.

To summarize, our model for the gluon field of a single nucleus deviates
in two ways from the McLerran-Venugopalan model. First, for simplicity we
use a mean-field approximation which reproduces 
the essential physics. Second, we implement a UV regulator $Q_0$.
To compare with the existing literature one can compute some quantities
in the limit $Q_0 \to \infty$. For the correlation function of charges 
one obtains 
$\langle \rho^a_k (\mathbf{x}_\perp) \rho^b_k (\mathbf{y}_\perp)\rangle
= \delta^{ab} (\sigma_k/6) \, \delta^2 (\mathbf{x}_\perp - \mathbf{y}_\perp)$
for constant densities $\sigma_k$ (here $a,b = 1,\ldots, N_c^2-1$)
\cite{MV,KMW,JMKMLW:96,Lappi:06}. In the same limit we find that the 
field correlator $\langle A^i(\mathbf{x}_\perp) A^i(\mathbf{y}_\perp)\rangle$
is a good approximation of the analytic form \cite{JMKMLW:96}.

We are now ready to use our model of the single nucleus gluon field
to obtain the initial electric and magnetic energy densities
$\epsilon_E = \langle \tr (F^{+-}_{(0)})^2 \rangle$ and
$\epsilon_M = \langle \tr (F^{21}_{(0)})^2 \rangle$.
After evaluating the expectation values
 $ \left\langle i^2 \tr \left([T_{1,u},T_{2,v}][T_{1,u'},T_{2,v'}]\right)
  \right\rangle
  = \delta_{uu'} \delta_{vv'} {N_{1,u} N_{2,v}}/{N_c}$
(the trace refers to color) we find
\begin{multline}
  \epsilon_{E}(\mathbf{x}_\perp)
  = \frac{g^6}{N_c} \sum_{u,v} N_{1,u} N_{2,v}
    \left(
  (\mathbf{x}_\perp - \mathbf{b}_{u})\cdot
  (\mathbf{x}_\perp - \mathbf{b}_{v})\right)^2 \\
  \times
  \frac{G(|\mathbf{x}_\perp - \mathbf{b}_{u}|)^2}{
  |\mathbf{x}_\perp - \mathbf{b}_{u}|^2}
  \frac{G(|\mathbf{x}_\perp - \mathbf{b}_{v}|)^2}{
  |\mathbf{x}_\perp - \mathbf{b}_{v}|^2}
  \, .
  \label{eq:edens}
\end{multline}
For $\epsilon_M$ the square on the first line has to be replaced by
$\left(\epsilon^{ij} \left(b^i_{u} b_{v}^j - x_\perp^i (b_{v}^j -b_{u}^j)
\right)\right)^2$.
We evaluate this result for the center ($\mathbf{x}_\perp=0$) of
two large nuclei with radius $R_A$ colliding head-on, so that
$R_A \gg R_c \gg \lambda$. The contributions of the two nuclei to
Eq.\ \eqref{eq:edens} factorize if the size of each cell is small,
as was also noticed in \cite{Lappi:06}.
Assuming that the charge densities $\sigma_k$ are roughly constant
in the center of each nucleus we find to good approximation that
\begin{equation}
  \epsilon_E  = \epsilon_M = \frac{1}{2} \epsilon_0 =
  \frac{\pi \alpha_s^3}{N_c} \sigma_1\sigma_2
  \ln^2 \left( 1+ c\zeta^2 \right)\, .
\end{equation}
This result only depends on the ratio of scales
$\zeta = R_c/\lambda$ and $c \approx 0.42$ is a numerical constant.

The charge densities $\sigma_k$, whose fluctuations create the color
distributions $\rho_k$, are given by the large-$x$ partons in the
nuclei.
To give a numerical estimate for two Au nuclei colliding at RHIC energy
we count all partons in the nuclei above the cutoff scale $Q_0$, similar
to the procedure in \cite{GyMLa:97}. In practice
we determine $\sigma=\sigma_1=\sigma_2$ as a function of $Q_0$ using CTEQ
parton distributions. Note that while the screening length $R_c$ in a nucleus
is a physical quantity, the cutoff $Q_0$ is unphysical. We observe that 
$Q_s =\sqrt{\alpha_s \sigma} \sim 1/R_c$ is indeed almost independent 
of $Q_0$; however, the energy density $\epsilon_0$ is not. The residual
logarithmic dependence on $Q_0$ should vanish if we match classical
and hard perturbative results in the region where they are comparable
\cite{GyMLa:97}. 

\begin{figure}[t]
\includegraphics[width=\columnwidth]{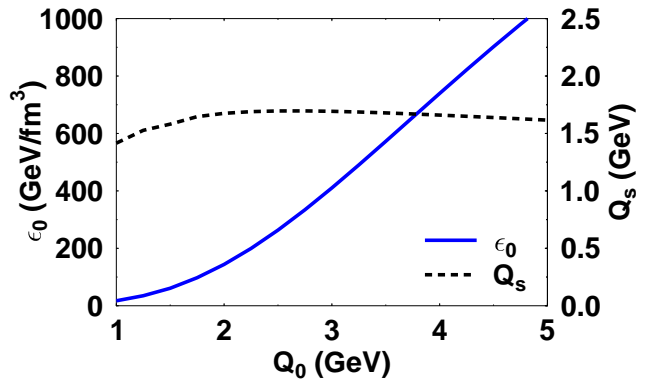}
\caption{\label{fig:1} Initial energy density $\epsilon_0$ for $\tau \to 0$
and saturation scale $Q_s$ at RHIC 
as a function of the UV cutoff $Q_0$.}
\end{figure}

Fig.\ \ref{fig:1} shows our estimate for the inital energy density
$\epsilon_0$ in the center of the collision as a function of the UV cutoff
$Q_0$ for central collisions at RHIC using $\zeta^2 = Q_0^2/(\alpha_s
\sigma)$. We find that $Q_s$ only varies between 1.4 and 1.7 GeV if $Q_0$
is varied between 1 and 10 GeV. For a reasonable cutoff $Q_0 = 2.5$ GeV we have
$\epsilon_0 \approx 260$ GeV/fm$^3$. This is compatible with
a value of 130 GeV/fm$^3$ at $\tau = 0.1$ fm/$c$ found by T.\ Lappi in 
\cite{Lappi:06}.
Note that $\epsilon_0$ only takes into account gluon modes with
transverse momentum less than $Q_0$. Results for finite
$\tau$ and the matching with hard processes to compute the total energy 
density and to eliminate the sensitivity to $Q_0$ will be 
discussed in a forthcoming publication \cite{FKLMB}.

To conclude, we introduced a near-field expansion to solve the
classical Yang-Mills equations for the collision of two large nuclei in
the color glass picture. We found that strong longitudinal chromoelectric
and magnetic fields dominate at early times. Using a
coarse--graining of color charges we derived a simple expression for the
initial energy density of the soft gluon field. A rough estimate
implies values of about 260 GeV/fm$^3$ at $\tau = 0$
for the center of two colliding nuclei at RHIC.

We would like to thank B.\ M\"uller and S.\ A.\ Bass for discussions, and
L.\ McLerran for his encouragement.
We are grateful to S.\ Jeon, D.\ Kharzeev, T.\ Lappi and R.\
Venugopalan for helpful conversations. This work was supported by
DOE grants DE-FG02-87ER40328, DE-AC02-98CH10886 and RIKEN BNL.

\end{document}